# Identifying Student Difficulties with Understanding Induced EMF

Scott Secrest[1,2] and Ingrid Novodvorsky[1]

[1]Dept. of Physics, University of Arizona, Tucson, AZ, USA
[2] St. Francis de Sales High School, Toledo, OH, USA
*(Note: Scott Secrest can be reached at ssecrest@sfstoledo.org)*

## Abstract

This paper reports on a study of student understanding of induced EMF. Students enrolled in a physics lab course completed a pretest and posttest that included questions about induced EMF. Following the posttest, some of those students participated in interviews in which they elaborated on their ideas. Pretest and posttest data are presented in the article, as well as sample interview text. After instruction in the lab course, most students were able to identify that moving a magnet near a coil would induce a current. However, nearly half were unable to correctly determine the direction of the induced current. Many students also demonstrated difficulty with the idea that an EMF is induced by a changing magnetic *flux*, not just a changing magnetic *field*.

## Introduction

Electricity and magnetism is an area in introductory physics that poses problems for both students and teachers. Students find the material challenging because it is difficult to visualize what is happening, and teachers find it challenging to teach because it is an abstract topic[1]. Often the course is taught with a heavy emphasis on formulas and solving quantitative problems, and with a lack of focus on conceptual understanding of the phenomena[2]. However, many research studies have suggested that students' ability to solve quantitative problems does not necessarily indicate conceptual understanding[3,4].

Another difficulty with teaching electricity and magnetism is students' lack of prior concrete experiences with the physical phenomena being described[5]. Because many different abstract concepts are presented in a short period of time, students have trouble absorbing and distinguishing among these concepts, which causes them to have incorrect pictures of the physical world. Students incorrectly state that an electric field will not have an effect on a stationary charge or that a magnetic field will exert a force on a charge even if it is stationary[5,6,7]. They confuse the electric and magnetic forces, which makes it difficult to understand induced EMF, or how a dipole antenna works[8].

The purpose of this paper is to document student difficulties with induced EMF that were observed with students enrolled in introductory physics at a large southwestern university. These observations were made as part of a larger study on the effectiveness of the introductory labs in electricity and magnetism for non-physics majors at the university. Documentation of these difficulties was the first step in addressing them effectively in both the lecture and in laboratory settings.

## Methods of Investigation

This study involved students enrolled in three sections of a non-majors lab course that includes topics in electricity and magnetism (seven labs) and optics (five labs). Most students in the lab are concurrently enrolled in either the calculus- or non-calculus-based, second-semester physics classes (which cover electricity and magnetism, optics, and some modern physics), but it is not required that students take both the lab and the lecture course at the same time.

In order to study what difficulties students had, even after induced EMF was addressed in both their lecture and lab classes, two methods were used. A pretest was given to students on the first day of lab and the same test was given again after the electricity and magnetism labs were completed. A total of 47 students took both the pretest and posttest. In addition, 30 individual student interviews were conducted, after students completed the posttest, to gain a better understanding of students' thinking.

### Pretest and Posttest

This test was made up of 27 multiple-choice questions that covered many of the important concepts in electricity and magnetism, including those related to circuits, induced currents, and electromagnetic waves. Five of the questions dealt with the idea of induced EMF and induced currents. Two of the questions were taken from *Peer Instruction: A Users Manual*, written by Eric Mazur[9]. The other three were general questions dealing with induced current and its direction. (The questions are included in the Appendix.)

The first question on induced EMF asked about a loop of wire whose area was increased in the presence of a uniform magnetic field. Students were asked if a current was induced in the loop, and if so to determine if the current was clockwise or counterclockwise. The second question involved a battery connected to a solenoid, with a bulb connected across the solenoid. Students were asked what happened to the light bulb when the switch was opened.

In the last three questions students were shown a picture of a bar magnet and a copper ring and asked if a current would be induced in the ring in three situations, holding the magnet stationary near the coil, moving the north end of the magnet away from the coil, and holding the magnet stationary further away from the coil. If a current would be induced, students were also asked to determine its direction.

### Interviews

To probe the students' understanding of concepts in electricity and magnetism that were taught in the lab, individual interviews were conducted with 30 students. The interviews were audiotaped and transcribed, and the transcripts were analyzed in detail. The interviews provided a better look at what difficulties the students still had after completing the lab course, and an opportunity to ask the students to explain their answers in detail.

The interviews included questions from the three major areas that were addressed in the labs: electric circuits, induced EMF, and electromagnetic waves, but only the section on induced EMF will be discussed in this paper. Demonstrations similar to those done in the lab course were set up, and the students were asked to explain what would be observed. Students were not told if their answers were correct during the interview because many of the questions were based on the same concepts. However, if an incorrect answer was given, other questions were asked to try to clarify the difficulty in understanding.

For the questions on induced EMF, various arrangements of a bar magnet and a coil of wire were presented, and students were asked if current was induced in the coil and in what direction it flowed. The situations involved moving the magnet along the axis of the coil and toward the side of the coil, holding the magnet stationary and rotating the coil, and shrinking a loop of wire with a magnet inside.

## Results and Discussion

Induced EMF is a topic about which students do not have many preconceived ideas, because it is something that most students have not experienced. However, current research still identifies difficulties with the topic[2]. In this study, questions 21-25 on the pretest and posttest examined whether students understood what circumstances cause a current to be induced in a circuit, and what determines the magnitude and direction of the induced current. Table 1 summarizes student responses on the pretest and posttest.

**Table 1**

| Question 21 | A* | B | C | | | Gain Score |
|---|---|---|---|---|---|---|
| Pretest | 40 | 36 | 24 | | | |
| | | | | | | 0.15 |
| Posttest | 49 | 49 | 2 | | | |
| Question 22 | A | B | C | D | E* | |
| | | | | | | -0.01 |
| Pretest | 6 | 25 | 39 | 14 | 16 | |
| Posttest | 22 | 13 | 39 | 11 | 15 | |
| Question 23 | A | B | C* | | | |
| | | | | | | 0.57 |
| Pretest | 38 | 36 | 26 | | | |
| Posttest | 21 | 11 | 68 | | | |
| Question 24 | A | B* | C | | | |
| | | | | | | 0.14 |
| Pretest | 50 | 36 | 14 | | | |
| Posttest | 36 | 45 | 19 | | | |
| Question 25 | A | B | C* | | | |
| | | | | | | 0.61 |
| Pretest | 24 | 48 | 28 | | | |
| Posttest | 13 | 15 | 72 | | | |

Percentage of students selecting each answer choice for questions dealing with induced EMF. The correct choices are indicated with asterisks. The gain scores are calculated by using $\frac{posttest\ \%correct - pretest\ \%correct}{100 - pretest\ \%correct}$.

The responses on the pretest were scattered among the different choices, showing that students did not hold just one incorrect idea. It appears that they were guessing from among the choices on most of the questions because roughly equal numbers of students chose each response, except for Question 22. This question had scattered results, but they were not equally divided among the choices.

From the posttest results it appears that most students learned that, if both the magnet and coil are stationary, there will be no current induced in the coil. Questions 23 and 25 addressed this idea; students were asked if there would be an induced current if a magnet were held stationary at two points, A and B. In Question 23 the magnet was held at point A, which was close to the coil. Most students (68%) stated that there would be no current induced in the coil. For Question 25, an even larger percentage of students (72%) gave the correct answer that no current would be induced if the magnet were held stationary at point B, possibly because some students felt that the magnet was not "strong enough" to induce a current farther away. This idea was expressed by a small number of students for similar questions in the interviews.

In the interviews students were asked if a current would be induced if a magnet were held stationary next to a coil. Most students correctly stated that there would not be a current flow. In addition, most students understood that if a magnet were moved toward a coil faster, the induced current in that coil would be larger. They also knew that if the strength of the magnetic field were increased, the induced current would increase as well. Other students gave the correct answers to these questions, but their reasoning was not as clear. The following student correctly answered the questions, but was not able to explain why she gave the answers she did. (SS: Interviewer, S 30: Student #30)

> SS:    Okay. The next question…you had said that moving a magnet towards the coil will induce a current to flow. If I move the magnet slowly towards the coil or if I move the magnet quickly towards the coil and I measure the current in those two cases, is the size of the current the same in those two cases?
> S30:    I don't think so
> SS:    Okay which one would be larger?
> S30:    When you move it faster.
> SS:    And why?
> S30:    Because there's just more movement, more I don't know, just more movement…like faster.
> SS:    Okay. Now if I move one magnet towards the coil and I measure the current that flows or if I move two magnets towards the coil and I measure the current that flows, is the size of the current the same in those two cases?
> S30:    I wouldn't think so. I would think the two…I would think that it would be more just because there's two.

This example is more typical of the responses that were given. Many students had difficulty explaining their reasoning, or the reasoning that they gave was not completely correct, but they remembered what they had seen in lab.

One concept that many students had difficulty with was determining the direction that induced current would flow in different situations. They could correctly identify that current would flow if a magnet were moved toward or away from a coil, but they could not determine the direction. The answers given to Question 24 on the posttest highlight this idea. In this question, the north end of a magnet was pointed toward a metal ring, and the magnet was moved away from the ring. Most of the students (81%) identified that a current would flow in the ring, but the students had difficulty determining the direction. More students gave the correct response, 45% versus 36%, but in two out of the three classes answers were evenly split between the two directions. (Determining the direction of the induced current was addressed in the labs.)

It was possible to explore this difficulty more in the interviews. Most students did not seem to recognize that the magnetic field of the induced current needed to be in the direction that would offset the change in magnetic flux. Many knew that they were "supposed to use the right hand rule" to determine the direction of the current, but they were unsure of how to do this or they applied the rule incorrectly.

> SS:     Okay, so you're trying to use a right hand rule and you seemed a little unsure exactly how to apply it. So just tell me, explain it again how you applied it just to get that direction for the current. So what, what was the right hand rule that you used?
>
> S19:    Alright, well, I'm using, uhh, let's see. Well, I'm using my fingers as the magnetic field, and then pointing that in the direction of the north side of the magnet because if I remember right, the magnetic field is supposed to come out of the north side and coil around to the south. So I'm pointing it toward the north side and uhh, I don't know, since the umm current is supposed to be perpendicular to that, I'm going to say that my thumb points around in the direction of the current.

This student's response demonstrates two difficulties in determining the direction of the induced current. First of all, he did not apply Lenz's Law to determine the direction of the induced magnetic field. Secondly, in applying the right-hand rule, he had switched what the fingers and thumb represent. Instead of the thumb pointing in the direction of the induced magnetic field and the fingers curling in the direction of the current in the coil, the student had reversed them.

Many students did not remember the right hand rule, or how to use it, and just guessed on the direction of the current for the first situation, and based the direction for other situations on their first answer. This allowed an opportunity to uncover some of the difficulties that students had with their reasoning, without considering their misuse of the right hand rule. For example, some only took the polarization of the magnet into account. In the interviews some students stated that if the north end of the magnet is pointing toward the loop the current is always clockwise, and if the south end is pointing toward the loop the current is always counterclockwise. The direction of motion of the coil did not matter. Other students only took the direction of movement into account. If the magnet was moving toward the coil, the current was clockwise, and if the magnet was moving away from the coil, the current was clockwise. It did not matter which end of the magnet was moving toward or away from the coil. It was only the direction of motion that determined the direction of the current.

A third topic that most students had difficulty with was the fact that changing the area of a loop in a magnetic field induced a current. This suggests that they did not completely understand what magnetic flux was, or that it is a changing magnetic *flux* that induces a current. They could state that a changing magnetic field induced a current, but did not realize that this was because changing the magnetic field meant that the magnetic flux was changing.

Question 21 on the posttest investigated this idea, picturing a loop of wire whose area was increased in the presence of a uniform magnetic field. Students were asked to determine if there was an induced current, and if so, the direction that it flowed. Only one student out of 47 (2%) stated that no current was induced, which would indicate that almost all students understood that changing the area of a loop did induce a current. In the interviews a similar question was asked. A magnet was held stationary, while the radius of a loop of wire around the magnet was decreased. Again, most students stated that a current was induced in the loop. However, only a few correctly stated that it was because the area was being decreased. The other responses were similar to the following:

> SS:    And so now I'm going to have the north end of the magnet inside the loop of wire, and then–if you just hold the magnet stationary. And I shrink down this loop. While I'm shrinking it down, does current flow in the loop? It's going from a large loop to a small loop, and as I'm doing that, does current flow?
> S9:    Yeah.
> SS:    And so why does current flow in this case?
> S9:    Umm...just because the forces acting on the coil will be greater when it's closer than when it's further apart.

Many students stated that because the loop was shrinking and the wire was coming closer to the magnet, the "force due to the magnetic field" was getting stronger, or that the "magnet is going to have a higher effect on the charges because it is closer." Others stated that it was because the "magnetic field at the loop" was getting stronger, trying to relate this to a changing magnetic field inducing a current.

Another interview question addressed the idea that a changing magnetic flux induces a current. A magnet was held near a coil of wire and the coil was rotated. The students were again asked to determine if there was an induced current and the direction that it flowed. For this situation, few students answered that a current would be induced. Of those, only one correctly stated that it was because the cross-sectional area was changing. The others stated that it was because the loop was moving, again only relying on motion to determine if current is induced. The following interview illustrates this idea.

> SS:    Okay, the next question is, I hold the north end of the magnet pointing towards the coil and I rotate the coil. So as I'm rotating the coil, does current flow?
> S21:   It would very minimally, cause as you rotate it the sides of the coils are moving towards the magnet so it would sort of be like moving the magnet towards the center of the coil, it would just not be very much because it's not much movement.

These responses show that students did not have as strong an understanding as the posttest might have indicated of why changing the area of a loop in a magnetic field induced a current. This difficulty has been corroborated by other studies[2].

**Conclusions and Recommendations for Further Research**

Most students were able to identify that moving a magnet near a coil would induce a current in the coil, and that, if both the magnet and coil are stationary, no current would flow. However, the students had trouble determining the direction of the induced current, and had difficulty identifying that a current would be induced in a loop rotating in a magnetic field. This implies that the students did not fully understand the concept of magnetic flux and its role in induced EMF. It is clear that more attention needs to be devoted to guiding students in the application of Lenz's law to determine the direction of the induced EMF, and in understanding the role of a changing magnetic flux in inducing an EMF.

Specifically, the lab activities that deal with these topics must guide students in the application of Lenz's Law to various situations, with the proper use of the right hand rule. Their predictions for the direction of induced current can then be tested in the lab, and they can be guided through the use of Socratic dialogue[10] to understand any deviations from their predictions. Similarly, students must encounter a variety of situations in which the magnetic flux, not just the magnetic field, is changing, make the same sort of predictions, test them in the lab, and then discuss the results.

These preliminary results can also provide a basis for further investigations in this area. Additional multiple-choice questions based on students' difficulties with induced EMF should be developed in order to continue this study with a larger group of students, and to assess the effectiveness of the lab activities described above.

## Appendix:  Pretest/Posttest Questions on Induced EMF

21.    In the following figure, the bar moves to the right with a velocity *v* in a uniform magnetic field that points out of the page.

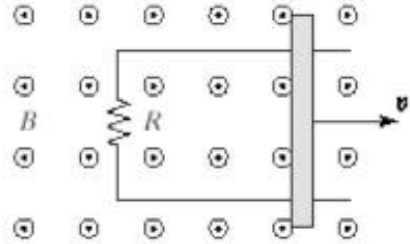

a.   The induced current is clockwise.
b.   The induced current is counterclockwise.
c.   There is no induced current.

22.    A battery is connected to a solenoid (long coil of wire) as shown below. When the switch is opened, the light bulb

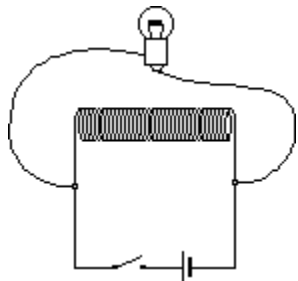

a.   remains off.
b.   instantaneously goes off.
c.   slowly dims out.
d.   keeps burning as brightly.
e.   flares up brightly, then dims and goes off.

In questions 23-25, refer to the following diagram.

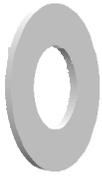     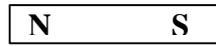          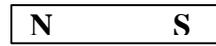

**Position A**                    **Position B**

23.    When the magnet is stationary at A, in which direction is the induced current in the loop?

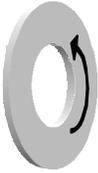                    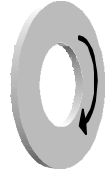

a.                                    b.

c.  There is no induced current.

24.    When the magnet is moving from A to B, in which direction is the induced current in the loop?

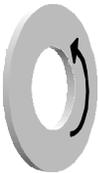                    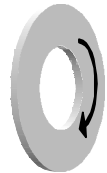

a.                                    b.

c.  There is no induced current.

25.    When the magnet is stationary at B, in which direction is the induced current in the loop?

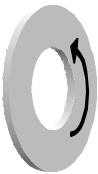                    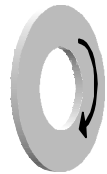

a.                                    b.

c.  There is no induced current.